\documentclass[final,3p,times]{elsarticle}
\usepackage{url,graphicx,graphics,amssymb,amsmath,relsize,supertabular}
\usepackage[pdfstartview={FitH},pdffitwindow=true,colorlinks=true,citecolor=blue,linkcolor=blue]{hyperref}%
\journal{Applied Radiation and Isotopes}
\begin{document}
\begin{frontmatter}
\title{Thickness Estimation of the Si Thin Films: a Simulation Study}
\author[dtp]{Mohammad Babazadeh}
\address[dtp]{Department of Theoretical Physics, Faculty of Physics, University of Tabriz, Tabriz,
Iran}
\author[damp]{Abdollah Rahmat Nezamabad}
\address[damp]{Department of Atomic and Molecular Physics, Faculty of Physics, University of Tabriz, Tabriz,
Iran}
\author[aa]{Hossein Movla\corref{cor1}}
\ead{h.movla@gmail.com Tel.:+98-9146352945} \cortext[cor1]{Corresponding author}
\address[aa]{Azar Aytash Co., Technology Incubator, University of Tabriz, Tabriz,
Iran}
\author[dtp]{Farzad Ghafari Jouneghani}
\begin{abstract}
We propose a theoretical study for Si thin film thickness measurement that is based on incident low energy electron beam on the film and counting the transmitted/incident electron fraction. It estimates the thin film thickness distribution from a exponential relation which obtained from counting the fraction of transmitted/incident electron at different thicknesses. By using this obtained equation, it is possible to estimate unknown thickness of the Si thin film. In order to calculate the Si thin film thickness estimation, the energy of the incident electron beams is varied from 6-12 $keV$, while the thickness of the Si film is varied between 100-400 $nm$. The most significant feature of this method is that no expensive instruments are required. As anticipated, the proposed method shows that there is a relationship between film thickness and incident beam energy, which by using this relationship, we can find unknown film thickness in 1-D and 2-D conditions. Other advantages include wide measurement range, no calibration need and simple method. Additionally, an investigation by different beam energies helps to avoid artefact from this method. All calculations were done by CASINO numerical simulation package.
\end{abstract}
\end{frontmatter}
\section{Introduction}
\label{int}
The study of the physical properties of thin films is important due to its multiple technological applications such as, modern optoelectronic devices, sensors, micro and nano electronic devices \cite{Zhang,Klauk,Movla2010,Rahmati}, and during recent years, new developments in materials synthesis and fabrication such as graphene, conducting polymers and Si-based nanomaterials have been created significant opportunities for thin-film materials and technology \cite{Novoselov,Brabec2003,Kelley}.\\ Scanning electron microscopy (SEM) is one of the most popularly used tools for thin films characterizing such as thickness measurement, grain boundary studies and etc. Recently, SEM and related instruments such as energy dispersive X-ray spectrometer (EDS) and Electron Spectroscopy for Chemical Analysis (ESCA) have attracted a lot of interests to research and application in material science and provide more information from bulk, thin film and coating samples \cite{Hufner,Goldstein,Strohmeier}.
\begin{figure}[h]
\begin{center}
\resizebox{0.33 \textwidth}{!}{
\includegraphics{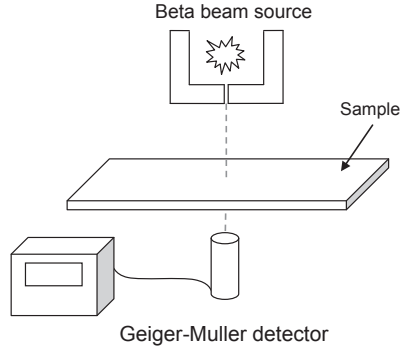}}
\end{center} \caption{Proposed experimental setup.}
\label{fig1}
\end{figure}
Materials mechanical, magnetic and electrical properties such as strength, conductance, permeability are all related to the thickness of the film. Therefore,  It is very important to measure the thickness of the films with high accuracy \cite{Pryds,EPFL,Jay}. On the other hand, variations in the thickness of the fabricated thin films have become important as the dimensions of the systems have been shrunk by device applications \cite{Pryds,Chopra,Movla2014-1}. Consequently, the thickness measurements have been become crucial for establishment of reliable thin film production. For different types and thicknesses of films, there are different methods for measurement. For example, if an optically transparent thin film on an opaque substrate is to be measured, one of the optical thickness measurement techniques such as ellipsometry can be used \cite{Jin}. If the sample is not optically transparent, and there are steps on the surface, the best way for measuring the thickness is scanning probe microscopy. Each measuring method has some advantages and disadvantages. But, the above mentioned techniques cannot be used in general labs and also, for a large area coated thin films, these techniques either require the use of an expensive camera or involve a complex procedure \cite{Kitagawa,Movla2014-2}. Most of today's available techniques are restricted to certain types of films and many have difficulties in performing the measurement in-situ \cite{Mann}. Measurement and estimation of the thin films thickness by using the simple and inexpensive techniques is too important parameter in both industrial and scientific aspects \cite{Kitagawa,Fathi}.
\\In this paper, we propose a theoretical study for Si thin film thickness measurement that is based on incident low energy electron beam on the film and counting the transmitted/incident electron fraction. It estimates the thin film thickness distribution from a exponential relation which obtained from counting the fraction of transmitted/incident electron at different thicknesses and by using the obtained equation, it is possible to estimate unknown thickness of the Si thin film. For the presented theoretical study, we propose a simple experimental set up which can be employed as experimental study.
\begin{figure}[h]
\begin{center}
\resizebox{0.4 \textwidth}{!}{
\includegraphics{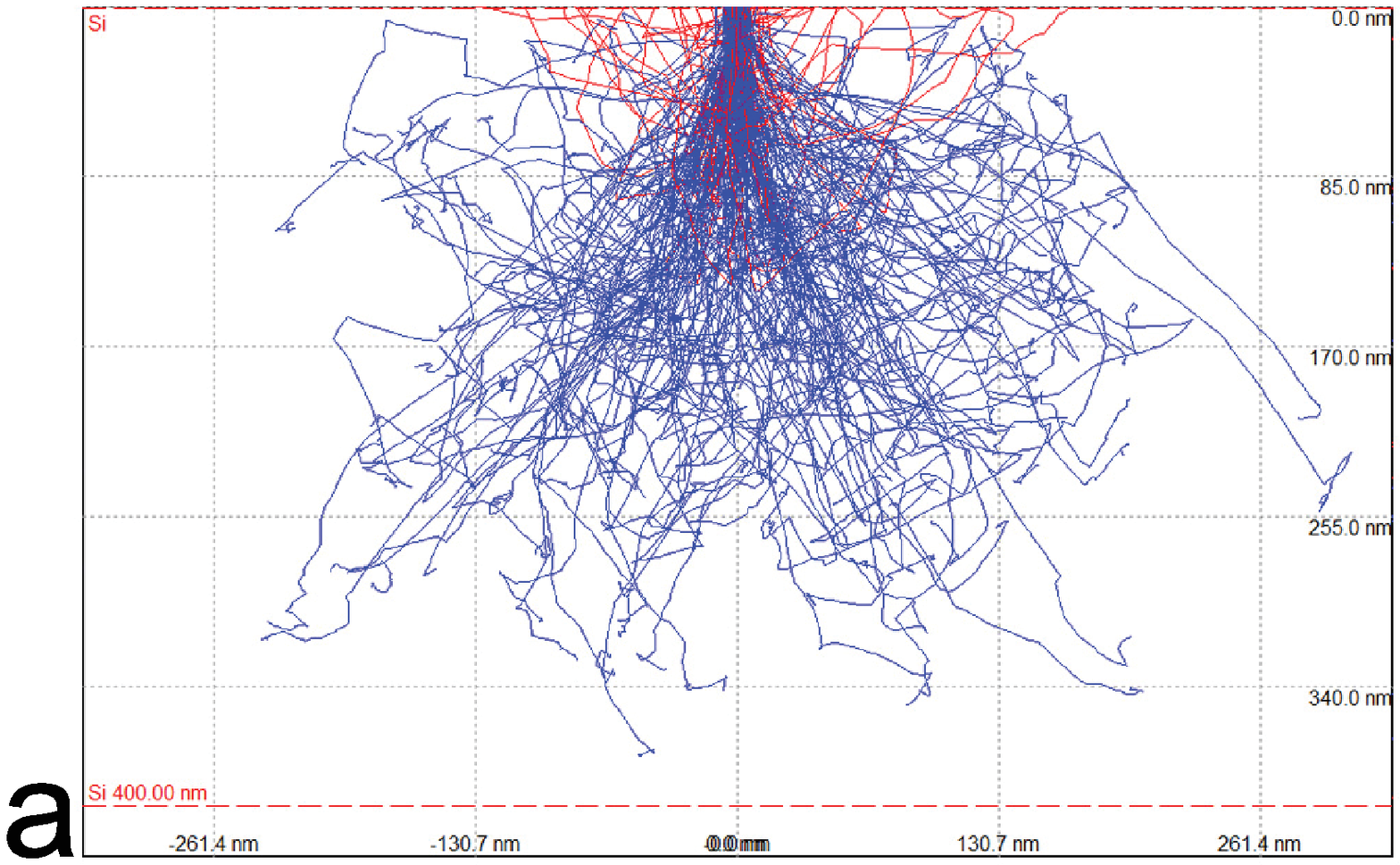}}
\resizebox{0.4 \textwidth}{!}{
\includegraphics{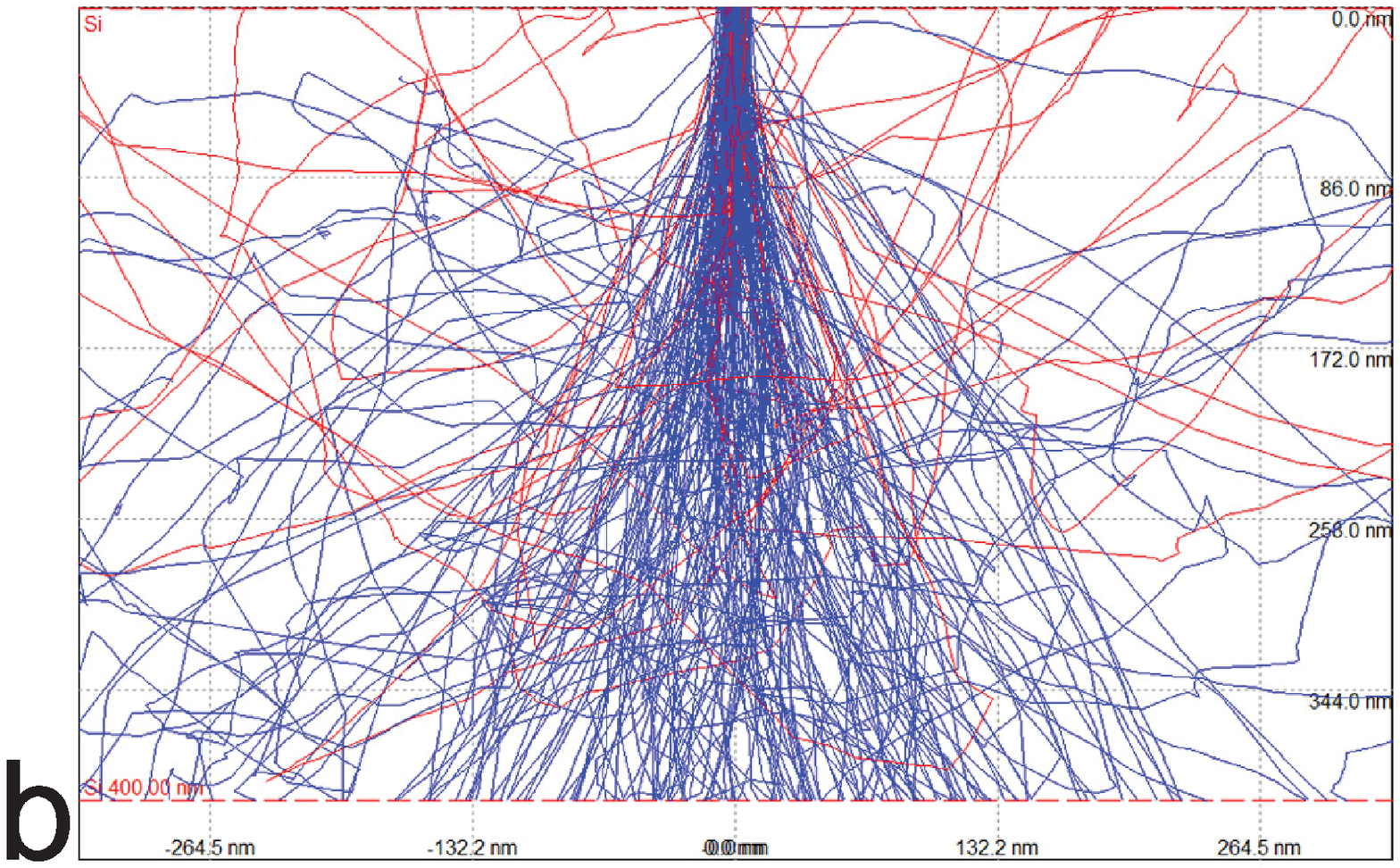}}
\end{center} \caption{Simulated trajectories of 50000 electrons in Si layer (400 nm) for (a) 6 keV and (b) 10keV incident electron beam energy. The trajectories have been projected into x--z plane. The red trajectory line represents the back-scattered electrons.} \label{fig2}
\end{figure}
As it is shown in Fig.~\ref{fig1}, by using a high energy electron source, such as nuclear beta-ray sources, and a Geiger-Muller detector, it is possible to count the number and the energy of the transmitted electrons. In different-thickness samples, because of the differences between received electrons by Geiger-Muller detector due to the sample thickness, it is possible to count transmitted versus incident electrons fraction.\\The most significant feature of this method is that no expensive instruments is required, also by adding more beam sources and detectors, it is possible to get a 2-D thickness measurement and estimation. Other advantages include wide measurement range, no calibration need, low cost and simple method. For simulation analysis, we used CASINO simulation software which developed by Raynald Gauvin et. al. at Universit\'{e} de Sherbrooke, Qu\'{e}bec, Canada \cite{Hovington1997,Drouin,Hovington}. In contrast to the perfect layer used in the our simulation, it should be noted that in the case of experimental use of this method, a calibration for Si thin film should be employed.
This program is a Monte Carlo simulation of electron trajectories in solids, specially designed to simulate the interaction of low energy electron beams with bulk samples and thin foils. A general description of how Monte Carlo calculations are
used to predict electron solid interactions can be found in a number of references \cite{Dapor,Joy,Gauvin1995,Gauvin2006,Demers}. The process involves calculating the trajectories of a large number of electrons striking a sample one at a time.
As it shown in Fig.~\ref{fig2}, the penetration depth of electrons into substrate material is higher for higher incident beam energy, according to this the using of different beam energy is one possibility of determining the thickness of the used materials.
The computation used tabulated Mott elastic scattering cross sections of Czyzewski and stopping powers model from Joy and Luo \cite{Drouin,Hovington}.
\section{Physical Model}
CASINO program is a single scattering Monte CArlo SImulation of electroN trajectory in sOlid specially designed for low-beam interaction in a bulk and thin foil \cite{Hovington1997}. In this section we describe the main routine that is used in this software. Firstly, a Gaussian distribution of the electron from the origin of the beam is used. The next step is to determine which atom is responsible for the elastic scattering. To achieve this, Eq.~\ref{Eq.1} is used \cite{Drouin}:
\begin{equation}
Random>\sum_{
i=1}^{n}\frac{\sigma_iF_i}{\sum_{
j=1}^{n}\sigma_iF_i},
\label{Eq.1}
\end{equation}
where "Random" is a random number uniformly distributed between 0 and 1, \ensuremath{\sigma_i} is the total cross section of element \ensuremath{i}, \ensuremath{F_i} is the atomic fraction of element \ensuremath{i}, and \ensuremath{n} is the number of elements in the region. \\When Eq.~\ref{Eq.1} is true, the responsible element for the collision is \ensuremath{i}. The polar angle of collision \ensuremath{\theta} is determined with the value of the partial cross section of the element \ensuremath{i}. This routine computes the polar angle of collision by solving:
\begin{equation}
R=\frac{\int_0^{\theta} \frac{d\sigma}{d\theta} sin(\theta)d\theta}
{\int_0^{\pi} \frac{d\sigma}{d\theta} sin(\theta)d\theta},
\label{Eq.2}
\end{equation}
where \ensuremath{\frac{d\sigma}{d\theta}} is the partial cross section and \ensuremath{R} is  a random number.
The azimuthal angle \ensuremath{\varphi} is uniformly distributed from o to \ensuremath{2\pi} and is given by Eq. ~\ref{Eq.3}:
\begin{equation}
\varphi=R\times2\pi,
\label{Eq.3}
\end{equation}
where \ensuremath{R} is another random number. The \ensuremath{\varphi} and \ensuremath{\theta} angles are defined as the angle formed by the last and new directions of the electrons, so we must recalculate the direction relative to a fixed axis.\\CASINO computes the direction \ensuremath{\cos (R_x,R_y,R_z)} with the old value \ensuremath{(R_{x0},R_{y0},R_{z0})}. \ensuremath{R_{x},R_{y},R_{z}} directions can be present by Eq.~\ref{Eq.4} to Eq.~\ref{Eq.6}, respectively  \cite{Drouin2007}.
\begin{align}
R_x &=\frac{R_{z0}sin\theta cos\varphi}{\sqrt{R_{x 0}^2\times R_{z 0}^2}}+ \nonumber\\
&\frac{R_{x 0}\times R_{y 0}\sin\theta \sin\varphi}{\sqrt{R_{x 0}^2\times R_{y 0}^2+(R_{x 0}^2\times R_{z 0}^2)\times(R_{x 0}^2\times R_{z 0}^2)+R_{y 0}^2\times R_{z 0}^2}}\nonumber\\
&+ R_{x 0}\cos\theta,
\label{Eq.4}
\end{align}
\begin{align}
R_y &=\frac{-R_{x 0}^2\times R_{z 0}^2\sin\theta \sin\varphi}{\sqrt{R_{x 0}^2\times R_{y 0}^2+(R_{x 0}^2\times R_{z 0}^2)\times(R_{x 0}^2\times R_{z 0}^2)+R_{y 0}^2\times R_{z 0}^2}}\nonumber\\
&+ R_{y 0}\cos\theta,
\label{Eq.5}
\end{align}
\begin{align}
R_z &=\frac{-R_{x 0}sin\theta cos\varphi}{\sqrt{R_{x 0}^2\times R_{z 0}^2}}+ \nonumber\\
&\frac{R_{z 0}\times R_{y 0}\sin\theta \sin\varphi}{\sqrt{R_{x 0}^2\times R_{y 0}^2+(R_{x 0}^2\times R_{z 0}^2)\times(R_{x 0}^2\times R_{z 0}^2)+R_{y 0}^2\times R_{z 0}^2}}\nonumber\\
&+ R_{z 0}\cos\theta,
\label{Eq.6}
\end{align}
We have noticed that their equation does not always satisfy the simple sum rule of cosine:
\begin{equation}
R_x^2+R_y^2+R_z^2=1,
\label{Eq.7}
\end{equation}
This equation must always be true to compute consistent direction.
To determine the distance (L) between two collisions, Eq.~\ref{Eq.8} is used:
\begin{equation}
L=\lambda log (RLPM),
\label{Eq.8}
\end{equation}
where $RLPM$ is a random number and \ensuremath{\lambda} is the electron mean free path. CASINO computes the electron mean free path using this equation:
\begin{equation}
\lambda=\frac{1\times10^{21}\sum_{i=1}^n\frac{C_iA_i}{\rho}}
{N_0\sum_{i=1}^nF_i\sigma_i} (nm),
\label{Eq.9}
\end{equation}
where \ensuremath{C_i} is the weight fraction, \ensuremath{F_i} is the atomic fraction, \ensuremath{A_i} is the atomic mass, and \ensuremath{\sigma_i} is the total cross section of element \ensuremath{i}. \ensuremath{\rho} is the density of the region and \ensuremath{N_0} is the Avogadro number. If \ensuremath{\sigma_i} is given in nm\ensuremath{^2}, the value of \ensuremath{\lambda} will be in nm as expected by the program.\\The energy lost during a travel distance , \ensuremath{L}, is a constant value, since a continuous slowing down approximation is used in this program. The energy at position \ensuremath{i} is computed by the following equation:
\begin{equation}
E_i=E_{i-1}+\frac{dE}{dS}L,
\label{Eq.10}
\end{equation}
where \ensuremath{E_{i-1}} and \ensuremath{E_i} are the respective energy at previous and current collision and \ensuremath{dE/dS} is the rate of energy loss. Before carrying out the next collision, the program will check whether the electron have escaped the region or the specimen.
\section{Results and Discussion}
\begin{figure}[h]
\begin{center}
\resizebox{0.38 \textwidth}{!}{
\includegraphics{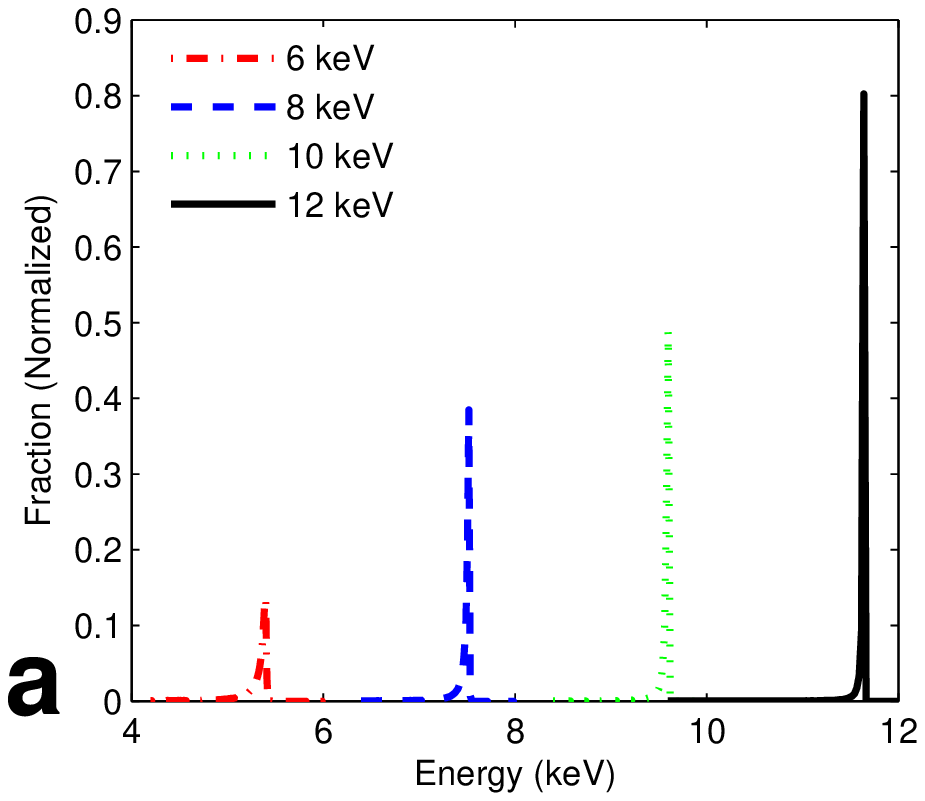}}
\resizebox{0.38 \textwidth}{!}{
\includegraphics{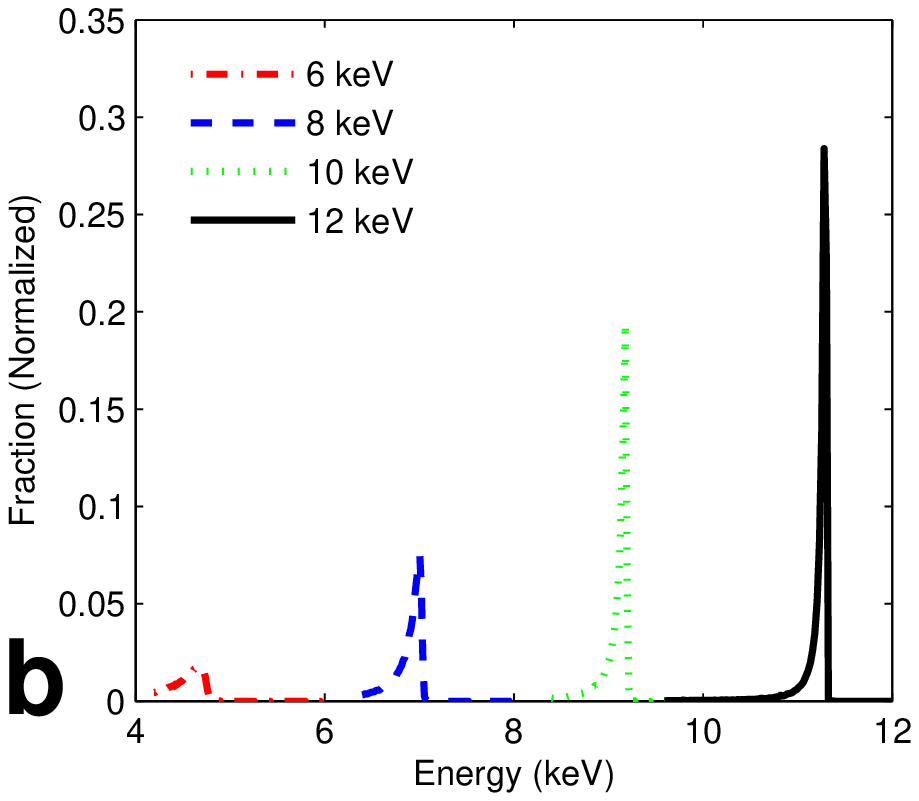}}
\resizebox{0.38 \textwidth}{!}{
\includegraphics{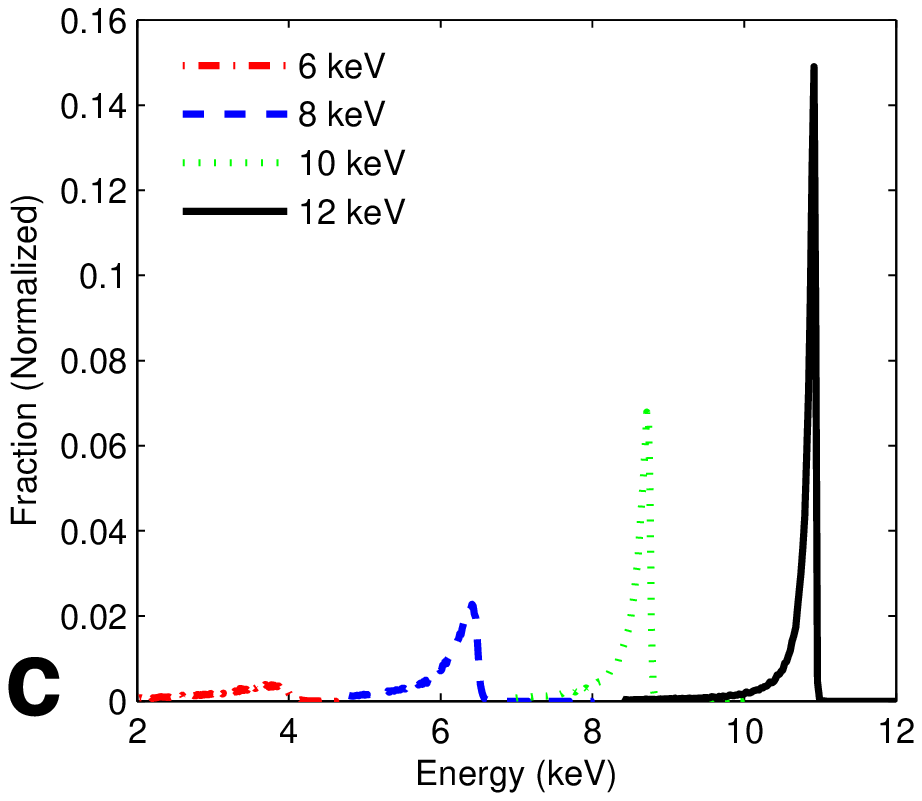}}
\resizebox{0.38 \textwidth}{!}{
\includegraphics{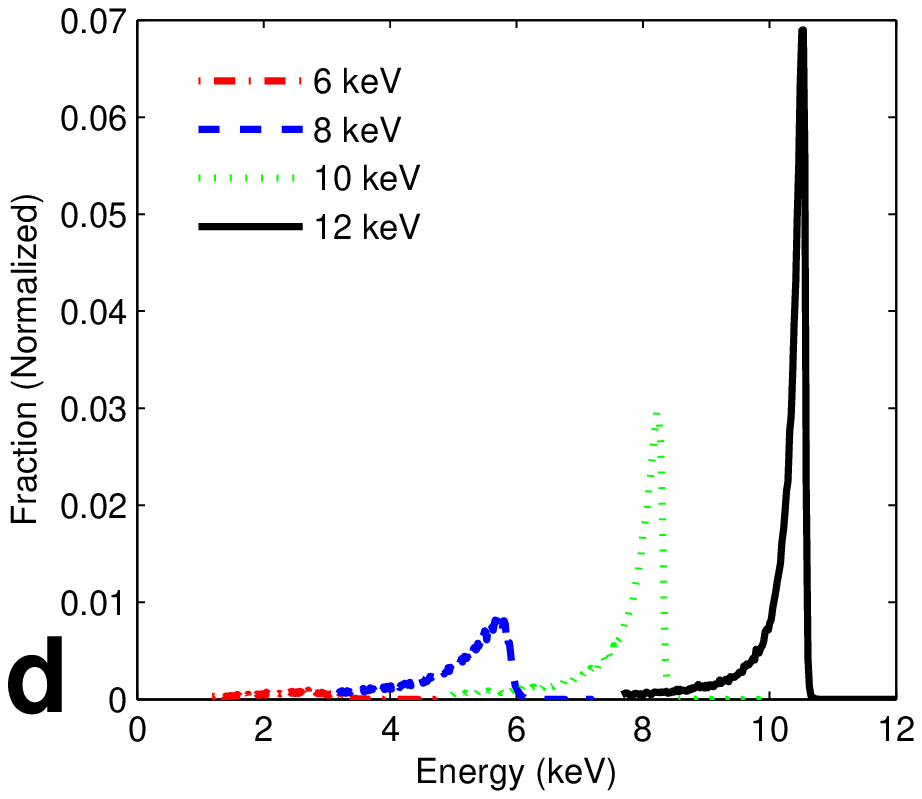}}
\end{center} \caption{Variation of the transmitted/incident electron fraction versus incident electron energy in (a) 100 nm, (b) 200 nm, (c) 300 nm, (d) 400 nm, respectively.}
\label{fig3}
\end{figure}
The energy of the incident electron beams is ranged from 6-12 keV, while the thickness of the Si films is between 100-400 nm. CASINO calculates the number and the energy of the transmitted electrons, while the incident energy and the film thickness are varied. Many cases, while changing the incident electron energy and the Si film thickness, have been simulated.
\begin{figure}[h]
\begin{center}
\resizebox{0.4 \textwidth}{!}{
\includegraphics{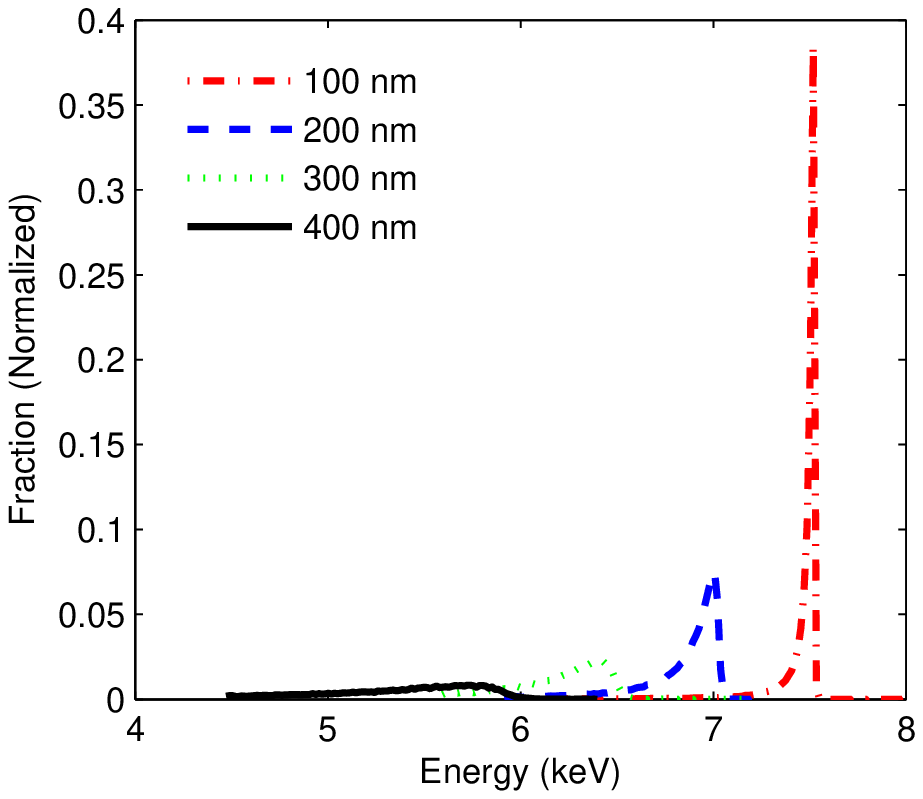}}
\end{center} \caption{Energy distribution of transmitted electrons at the different Si thickness at 8 keV.}
\label{fig4}
\end{figure}
\\Fig.~\ref{fig3}(a)-(d) show the fraction of the transmitted/incident electrons of the 100 nm to 400 nm Si samples at 6, 8, 10 and 12 keV, respectively. In this figures, it is shown that by increasing the incident electron beam energy, fraction of the passing electrons increases and due to this fact that at higher energy beams, large part of the incident electron can be transmitted from the Si samples. In fact the penetration depth of electrons into substrate material is higher for higher incident beam energy and leads to much received electrons at the other side of the sample which can be detect by electron detectors. In the 100 nm sample, Fig.~\ref{fig3}(a), It is shown that in 6 keV, 85 $\%$ of the incident electrons absorbed and only 15 $\%$ can be transmitted Si sample, but at 12 keV, about 80 $\%$ of the incident electrons transmit through the sample, which is about five times more than 6 keV incident electron beam.
\begin{figure}[h]
\begin{center}
\resizebox{0.4 \textwidth}{!}{
\includegraphics{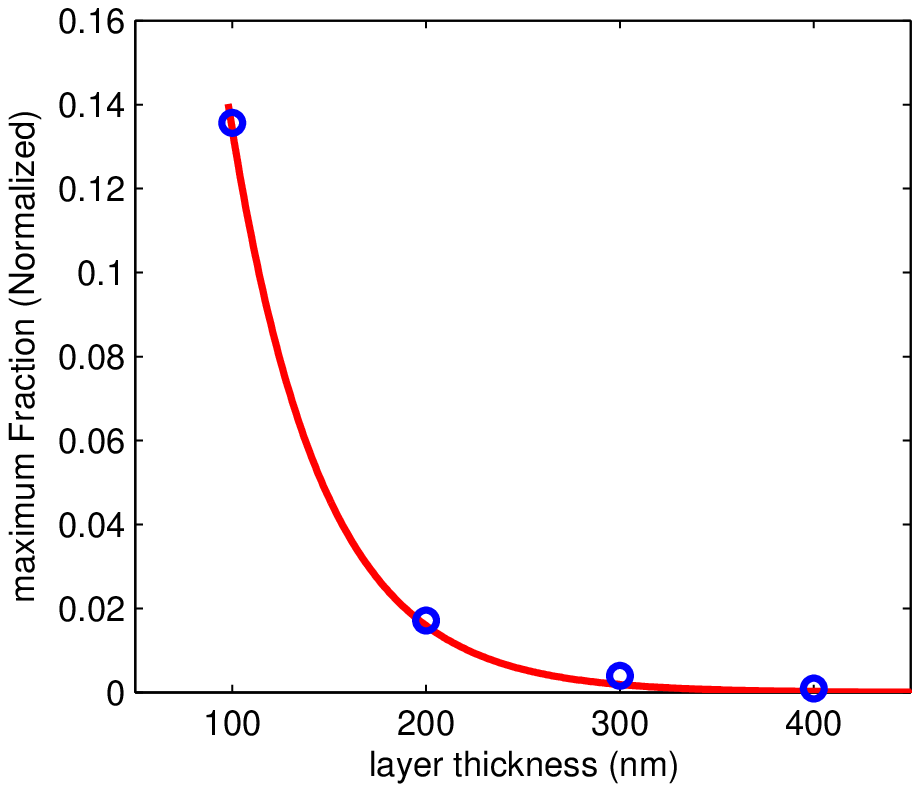}}
\end{center} \caption{Exponential fit of the maximum transmitted/incident fraction versus 100 nm Si layer thickness at 6 keV.}
\label{fig5}
\end{figure}
\begin{figure}[h]
\begin{center}
\resizebox{0.4 \textwidth}{!}{
\includegraphics{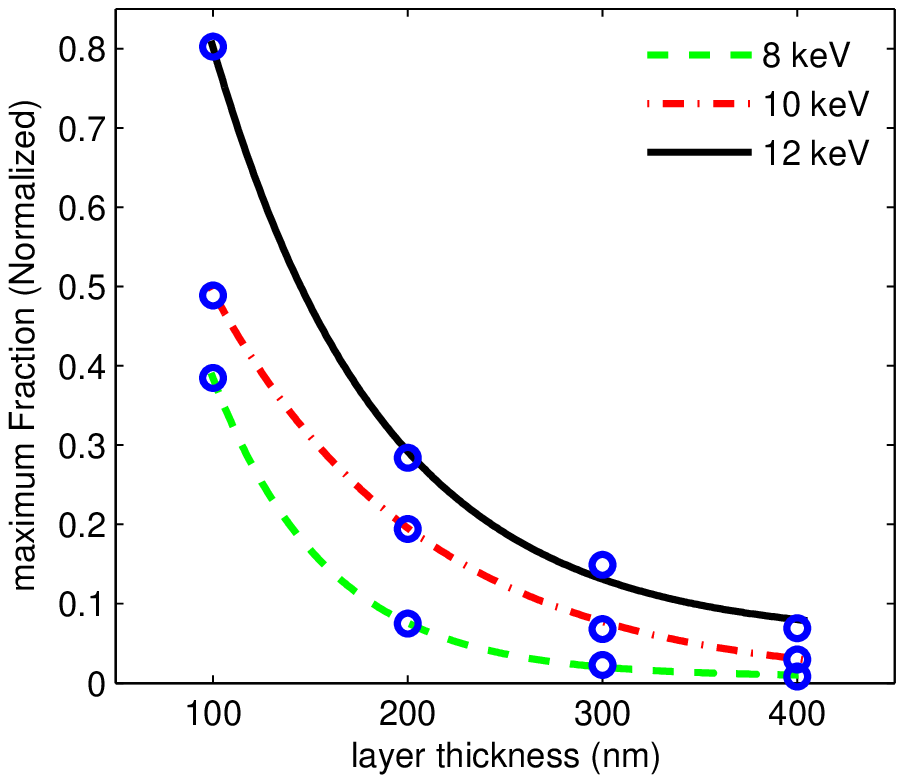}}
\end{center} \caption{Exponential fit of the maximum transmitted/incident fraction versus 200 nm, 300 nm and 400 nm Si layer thickness at 8, 10 and 12 keV.}
\label{fig6}
\end{figure}
As we can see from Fig.~\ref{fig3}(d), the electron beam with 6 keV energy cannot transmitted from the Si layer with 400 nm thickness. In other words, all the electrons of this beam are absorbed by the Si layer. By increasing the energy of beam, we can see that the number of electrons which are transmitted from the layer is increased. Moreover, by increasing the energy of beam, the energy of maximum fraction of transmitted electrons is increased and tends towards the initial energy of beam. By simulate these fractions, it is the first step to find a relationship between incident beam and sample thickness. \\Fig.~\ref{fig4} shows the number of the transmitted electrons through the 100-400 nm Si sample at the 8 keV energy of electron beam. It is shown in Fig.~\ref{fig4}, that by decreasing the Si thickness, number of the transmitted electrons increases.
Fig.~\ref{fig4} shows that when a 8 keV electron beam is applied, the peak energy of the transmitted electrons shifts from about 5.8 keV for a 400 nm Si film to about 7.5 keV for 100 nm film. In other words, for an electron beam with constant energy (8 Kev in Fig.~\ref{fig4}), by decreasing the thickness of a Si sample, the transmitted incident fraction is increased.
As it is shown in Fig.~\ref{fig4}, in the thicker layers, large part of the incident electrons absorbed and cannot penetrate through the sample.
For example, in 400 nm sample, maximum energy of the transmitted electrons is about 5.8 keV and only 0.2 $\%$ of the incident electron can pass the sample, but in 100 nm sample, energy range is about 7.5 keV and 60 $\%$ of the incident electrons is absorbed and 40 $\%$ is transmitted through the sample. According to simulation results, it was found that electron transmitting fraction in 400 nm sample is about twenty orders of magnitude less than 100 nm sample with the same incident electron beam energy. By using the simulation results for 100-400 nm Si samples, it is possible to estimate the Si sample thickness.
\\Fig.~\ref{fig5} shows the maximum transmitted/incident fraction versus Si layer thickness at 6 keV incident electron beam.
Simulation results show that maximum fraction versus Si layer thickness has an exponential relation. By exponential fitting the obtained data from maximum fraction versus thickness, this relation is given as:
\begin{equation}
T(x)=A_{E}exp(-\alpha_E x),
\label{Eq.10}
\end{equation}
Where x is the sample thickness in nm and $T(x)$ obtained as the fraction of transmitted/incident electrons. In the case of the 6 keV incident electron beams, $A_E=1.131$ and $\alpha_E=0.02131$ and $R^2=0.9984$. By using the above mentioned equation, it is possible to estimate unknown thickness of the Si layer at certain energy.
\begin{table}[h]
\begin{center}
\caption{The parameters of the Simulation}
\begin{tabular}{ccccc}
\hline\noalign{\smallskip}
Parameter &    6 keV   &     8 keV    &   10 keV  & 12 keV \\
\noalign{\smallskip}\hline\noalign{\smallskip}
$A_E$     & $1.131$    &   $2.094$    & $1.245$   &$2.354$ \\
$\alpha_E$& $0.0213$   &   $0.0175$   & $0.0093$  &$0.0115$ \\
\noalign{\smallskip}\hline
\end{tabular}\label{tab:1}
\end{center}
\end{table}
\\For the unknown film thickness, only we should count the number of received beta particles before putting sample in the front of detector and also after putting sample in above the detector at 6 keV. By calculating the fraction of counted number of beta particle at certain energy without and with sample, we can find transmitted/incident fraction, and by putting this fraction in Eq.~\ref{Eq.10}, we can find the thickness of the Si thin film sample.Fig.~\ref{fig6} shows the maximum transmitted/incident fraction versus Si layer thickness at 8, 10 and 12 keV incident electron beams. Same as the 6 keV case, we can obtain a exponential equation to determine the unknown film thickness in different incident beam energy with the different $A_E$ and $\alpha_E$ coefficients. These coefficients has listed in Tab.~\ref{tab:1}. \\By using this method it is possible to determine unknown film thickness of Si and also, it can be used for thickness estimation of other thin films. Additionally, an investigation by different beam energies helps to avoid artefact from this method. In our simulation, we assumed a Si thin films without any impurities, therefore, in experimental investigating with the different film thicknesses, it should be noted that there is a little difference between simulation method and obtained experimental data which can be caused by impurities of thin films.
\section{Conclusion}
We proposed an analysis method for determining the
thickness of Si thin films which can be further developed for other thin films. Numerical simulations have been done by adjusting parameters such as the electron stopping range, Si layer thickness, incident electron beam energy, and etc. In order to calculate the Si thin film thickness, the energy of the incident electron beams was varied from 6-12 $keV$, while the thickness of the Si film was varied between 100-400 $nm$. The thin film thickness distribution can be estimated by a exponential relation which obtained from counting the fraction of transmitted/incident electron at different thicknesses. This method has many advantages such as, wide measurement range, no calibration need, simplicity and low cost, so because of these advantages, this method can be employed at general labs and industrial purposes to thickness estimation and measurement of thin film sheets.
\section*{Acknowledgment}
The authors would like to express their thanks to Prof. Dr. Raynald Gauvin, Universit\'{e} de Sherbrooke, Qu\'{e}bec, Canada, for providing the CASINO simulation software.

\end{document}